\documentclass[11pt,preprint]{revtex4}
\usepackage{graphicx}
\usepackage{dcolumn}
\usepackage{bm}
\begin{document}

\title{Relativistic effects in neutron-deuteron elastic scattering}

\author{H.~Wita{\l}a}
\affiliation{M. Smoluchowski Institute of Physics, Jagiellonian
University,
                    PL-30059 Krak\'ow, Poland}

\author{J.~Golak}
\affiliation{M. Smoluchowski Institute of Physics, Jagiellonian
University,
                    PL-30059 Krak\'ow, Poland}

\author{W.\ Gl\"ockle}
\affiliation{Institut f\"ur theoretische Physik II,
Ruhr-Universit\"at Bochum, D-44780 Bochum, Germany}

\author{H.\ Kamada}
\affiliation{Department of Physics, Faculty of Engineering,
Kyushu Institute of Technology, Kitakyushu 804-8550, Japan}

\date{\today}

\begin{abstract}
 We  solved the  three-nucleon (3N) Faddeev
 equation including  relativistic features  at 
 incoming neutron lab energies  $E_n^{lab}=28$, $65$, $135$ and $250$~MeV.
Those features are relativistic kinematics, 
boost effects and Wigner spin rotations.
 As dynamical
 input a relativistic nucleon-nucleon (NN) interaction exactly on-shell
 equivalent to the AV18 NN potential has been used. The effects of
 Wigner rotations for elastic scattering observables were found to be
  small. The boost effects are significant  at higher energies.
They diminish the transition matrix elements at higher energies and lead
 in spite of the increased
 relativistic phase-space factor as compared to the nonrelativistic one to 
 rather small effects in the cross section, which are mostly  restricted
 to the backward angles. 
\end{abstract}


\maketitle
\setcounter{page}{1}

\section{Introduction}

High precision  nucleon-nucleon (NN) potentials such as
AV18~\cite{AV18}, CDBonn~\cite{CDBONN}, Nijm I,  II and 93~\cite{NIJMI}
describe the
NN data set up to about 350 MeV very well.
When these forces are used to predict binding energies of three-nucleon (3N)
systems they underestimate the experimental bindings of $^3H$ and $^3He$ by
about 0.5-1 MeV~\cite{Friar1993,Nogga1997}. This missing binding energy
can be cured by introducing a three-nucleon force (3NF) into the nuclear
Hamiltonian~\cite{Nogga1997}.

The study of elastic nucleon-deuteron (Nd) scattering
and nucleon induced deuteron breakup
 revealed a number of cases where the nonrelativistic description
based on pairwise forces only is insufficient to explain the data.
Generally, the studied
discrepancies between a theory based on NN potentials only  and experiment
  become larger  with increasing energy of the 3N system. Adding
now a  3NF to the  pairwise interactions  leads in some cases
to a better description of the data. The elastic Nd
angular distribution in the region of its minimum and at backward
angles is the best studied example~\cite{wit98,sek02}.
The clear discrepancy in this angular regions at 
energies below  $\approx100$~MeV 
nucleon lab energy between
a theory  based on  NN potentials only  and the cross
section data can be removed by
adding modern 3NFs to the nuclear Hamiltonian. Such a 3NF must be
adjusted with each  NN potential separately to
the experimental binding of $^3H$ and
$^3He$~\cite{wit98,wit01,sek02}.
 At energies higher than  $\approx 100$~MeV current
3NFs only partially improve the
description of cross section data and the remaining
 discrepancies, which  increase with energy,
 indicate the  possibility of relativistic effects.
The need for a  relativistic description of 3N scattering was also
raised  when precise
measurements of the total cross section for neutron-deuteron (nd)
scattering~\cite{abf98} were analyzed within the framework of
nonrelativistic Faddeev calculations~\cite{wit99}.
NN forces alone were
 insufficient to describe the data above  $\approx 100$~MeV.
 The effects due to relativistic kinematics considered in
 ref.~\cite{wit99}
were comparable at higher energies to the effects due to 3NFs.
 These indications show the
importance of a study taking  relativistic
effects in the 3N continuum into account.

The estimation of relativistic effects on the binding energy 
of three nucleons
has been the focus of a lot of work. Basically two different
aproaches have been followed: one is a
manifestly covariant scheme linked to a field theoretical
approach~\cite{Rupp92,Sam98,stad97}, the other
one is based on  relativistic quantum mechanics
formulated on spacelike hypersurfaces in Minkowski
space~\cite{bak52,foldy74,keister91,kei91}. Within the
second scheme the relativistic Hamiltonian  for on the mass shell particles
consists of relativistic
kinetic energies and two- and many-body interactions including their
boost corrections, which are dictated by the
Poincar\'e algebra~\cite{bak52,foldy74,keister91, kei91}. The applications of
these
two types of approaches to the 3N bound state have led to contradictory results. In the
approach based on field theory, relativistic effects increase the
triton binding energy, while in the approach based on relativistic
Hamiltonians they decrease the triton binding energy. This requires further insights which
is beyond the scope of this paper.

Due to the increased complexity of 3N scattering calculations as compared to
the bound state problem no results for
the 3N continuum including relativity are available. In order to extend the
Hamiltonian scheme in equal time formulation to 3N scattering one needs as a
starting point the Lorentz boosted NN potential which generates the NN
t matrix in a moving frame via a standard Lippmann-Schwinger equation. Such
potentials have been worked out and applied to a 3N bound state
in~\cite{kam2002}. The results obtained  supported the relativistic
effects found before in a relativistic quantum mechanics
approach~\cite{relform}.
The starting point for a NN potential in an arbitrary
moving frame is the interaction in the two-nucleon c.m. system, which enters
a relativistic NN Schr\"odinger or Lippmann-Schwinger equation. It differs from the
nonrelativistic Schr\"odinger equation just by the relativistic form for the kinetic energy.
 The
current realistic
NN potentials are defined and fitted in the context of the nonrelativistic
Schr\"odinger equation. Up to now NN potentials refitted with the same
accuracy in the framework of the relativistic NN Schr\"odinger equation do
not exist. In~\cite{kam2002} such refitting was omitted and
an analytical scale transformation of momenta which relates
NN potentials in the nonrelativistic and relativistic Schr\"odinger
equations in such a way, that exactly the same NN
phase shifts are obtained by both
equations, was employed~\cite{kam98}. 

Though this transformation is not a substitute for
a NN potential with proper relativistic features it can serve as a first step
to illustrate the effects of  Lorentz boosts on NN potentials. Such an
approach was applied in~\cite{kam2002} and we also will follow it in the
present study to get the first estimation of relativistic effects in the 3N
continuum.
 In this first study we would like  to find out what
 are the changes of elastic nd scattering observables
  when the  nonrelativistic form of the kinetic energy is replaced
by the relativistic one  and a  proper treatment of  boost effects and
 effects due to  Wigner
rotations of spin states is performed.

The paper is organized as follows. 
In Sec. II we lay out the relativistic features
underlying our treatment for a relativized Faddeev 
equation in the 3N continuum. This
incorporates the definition of the boosted two-body force, 
the various two-and three-body
states in general frames, the Wigner rotations and the 
singularity structure of the
relativistic free 3N propagator. Our manner to treat  
the 3N Faddeev equation is guided by
the lines presented in ~\cite{wit88} for the nonrelativistic case.  
In Sec. III we focus on the relativistic 
NN potential  and discuss the quality of
different approximations  for the boosted potential.
As a consequence, in this first study we  restrict ourselves to the
leading order relativistic  term in the expansion  of the boosted
potential.
In Sec. IV we apply our
formulation based on  a relativistic NN interaction which is exactly
 on-shell
equivalent to the nonrelativistic AV18 potential and
 solve the relativized
3N Faddeev equation with  different  approximations for the boost.
 We show and discuss results
for elastic Nd scattering.
 Sec. V contains a summary and outlook.

\section{Formulation}

The nucleon-deuteron scattering with neutron and protons interacting
through a NN potential $V$ alone  is described in terms of a breakup operator T
satisfying the Faddeev-type integral equation~\cite{wit88,glo96}

\begin{eqnarray}
T\vert \phi >  &=& t P \vert \phi > + t P G_0 T \vert \phi > .
\label{eq1a}
\end{eqnarray}
The two-nucleon (2N) t-matrix t results from the interaction $V$ through
the Lippmann-Schwinger equation. 
The permutation operator  $P=P_{12}P_{23} + P_{13}P_{23}$ is given in terms
of the transposition $P_{ij}$ which interchanges nucleons i and j. The incoming state  
$ \vert \phi > =
\vert \vec q_0 > \vert \phi_d > $ describes the free
nucleon-deuteron motion with relative momentum $\vec q_0$ and
the deuteron wave function $\vert \phi_d >$. Finally $G_0$ is the free 3N propagator. 
 The physical picture underlying Eq.(\ref{eq1a})  is
revealed after iteration which  leads to  a multiple scattering series 
 for T.  

The  elastic nd scattering transition operator U 
 is  given in terms of T by~\cite{wit88,glo96}
\begin{eqnarray}
U  &=& P G_0^{-1} + P T  .
\label{eq1c}
\end{eqnarray}


This is our standard nonrelativistic formulation, which is equivalent to the
nonrelativistic 3N Schr\"odinger equation plus boundary conditions.
The formal  structure of these equations  in the relativistic case remains
the same but the ingredients change. As explained in ~\cite{relform} 
the relativistic 3N Hamiltonian has the same form as the nonrelativistic one,
 only the momentum dependence of the kinetic energy changes and the relation
of the pair interactions to the ones in their corresponding c.m. 
frames changes,
too. Consequently all the formal steps 
leading to Eqs.(\ref{eq1a}) and (\ref{eq1c}) remain the same.

The relativistic kinetic energy of three equal 
mass nucleons  in their c.m. system can
conveniently be presented by introducing the free two-body mass operator.
Let $\vec k$ and $-\vec k$ be 
the momenta in one of the two-body subsystems, then
$2\omega({\vec k}) \equiv 2\sqrt{m^2 + {\vec k}^{\ 2}}$ is 
the momentum dependent 
2N mass operator
 and the 3N kinetic energy can be written as
\begin{eqnarray}
H_0 &=& \sqrt{(2\omega({\vec k}))^2 + {\vec q}^{\ 2}} 
+ \sqrt{m^2 + {\vec q}^{\ 2}} ,
\label{eq1an1}
\end{eqnarray}
where $\vec q$ is the momentum of the third 
particle and $-\vec q$ the total momentum of
the chosen two-body subsystem (m is the nucleon mass). Any two-body subsystem
can be chosen. As introduced in ~\cite{relform1} the pair forces in
the relativistic 3N Hamiltonian living in moving frames are chosen as
\begin{eqnarray}
V(\vec q \, ) \equiv \sqrt{( 2\omega(\vec k) + v)^2 + {\vec q}^{\ 2}} 
 - \sqrt{ (2\omega(\vec k))^2 + {\vec q}^{\ 2}} , 
\label{eq1}
\end{eqnarray}
where $ V(\vec q)$ 
for $\vec q = 0$
reduces to the potential $v$ defined in the 2N 
  c.m. system.  Note that also in that system the relativistic kinetic energy
of the two nucleons has to be chosen, which together with $v$ defines the 
interacting two-nucleon mass operator occuring in Eq.(\ref{eq1}). 

Let us now firstly regard the 2N subsystem. The standard nonrelativistic
2N Lippmann Schwinger equation turns now into a relativistic one, which
in a general frame reads
\begin{eqnarray}
 t( \vec k, \vec k~' ;  \vec q~) &=& 
 V( \vec k, \vec k~' ;  \vec q~) + 
\int d^3k'' { 
{V( \vec k, \vec k~'' ;  \vec q~)
 t( \vec k~'', \vec k~' ;  \vec q~) } 
\over{ \sqrt{ {( 2\omega( {\vec k}^{\ '} )^{\ 2} + {\vec q}^{\ 2}} }  
-  \sqrt{ {(2\omega(  {\vec k}^{\ ''})^{\ 2} 
+ {\vec q}^{\ 2}} } + i\epsilon  } } .
\label{eq2a}
\end{eqnarray}

We refer to $t(\vec k, \vec k~'; \vec q~)$ 
 as the boosted 2N t-matrix like we talk of
 the boosted  2N potential in Eq.(\ref{eq1}).

 Using (\ref{eq1}) the relativistic 2N Schr\"odinger equation for the deuteron
 in a moving frame 
 can be cast into the form
\begin{eqnarray}
 \phi_d( \vec k) &=& 
{ {1} \over { \sqrt{ {M_d^2 + {\vec q}^{\ 2}_0}} -  
\sqrt{ {( 2\omega (\vec k))^2 + {\vec q}^{\ 2}_0} } } } 
\int d^3k' 
 V( \vec k, \vec k~' ;  \vec q_0 ) 
 \phi_d( \vec k~') ,  
\label{eq2b}
\end{eqnarray}
where $\sqrt{ {M_d^2 + {\vec q}^{\ 2}_0}}$ is the energy of 
the deuteron in motion and
 $M_d$ its rest mass. This equation is a good check for the correct numerical
implementation of the boosted potential $V(\vec q)$  
as will be used below.

The new relativistic ingredients in Eqs.(\ref{eq1a}) 
and (\ref{eq1c}) will therefore  
be the boosted
 t-operator and the relativistic 3N propagator
\begin{eqnarray} 
G_0 &=& {{1}\over{E + i\epsilon - H_0}} ,
\label{eq2c}
\end{eqnarray}
where $H_0$ is given in Eq.~(3) and $E$ is the total 3N c.m. 
energy expressed in terms of the initial
neutron momentum $\vec q_0$ relative to the deuteron
\begin{eqnarray}
E &=& \sqrt{(M_d)^2 + {\vec q}_0^{\ 2} } + \sqrt{m^2 + {\vec q}_0^{\ 2}} .
\label{eq2d}
\end{eqnarray}

Currently Eq.(\ref{eq1a}) in its nonrelativistic form is numerically solved for
any NN interaction using a momentum  space partial wave decomposition.
Details are presented in ref.~\cite{wit88}. This turns Eq.(\ref{eq1a}) 
into a coupled set
of two- dimensional integral  equations.  As we show now, in the
relativistic case we can keep the same formal structure, though the
kinematics and the momentum representation of the permutation operator
 P is more complex and boosted t-operators as well as Wigner 
rotations will appear.

In the nonrelativistic case the partial wave projected momentum space basis is
\begin{eqnarray}
 &~& \vert p q (ls)j 
(\lambda{1\over{2}})IJ(t{1\over{2}})T > ,
\label{eq2e}
\end{eqnarray}
where p and q are the magnitudes of standard Jacobi momenta (see \cite{book}) 
and $(ls)j$ two-body quantum numbers with obvious meaning, 
$(\lambda 1/2)I$ refer to the third nucleon (described by the momentum q),
$J$ is the total 3N angular momentum and the rest are isospin quantum numbers.
This is now to be generalized to the relativistic case. 

We regard firstly
the two-nucleon system and replace the nonrelativistic relative 
two-nucleon momentum
$\vec p$ by $\vec k$, where $\vec k$ and 
$-\vec k$ are related to general momenta of two
nucleons, say $\vec p_2$ and $\vec p_3$,  
by a Lorentz boost:
\begin{eqnarray}
\vec k &\equiv& \vec k(\vec p_2, 
\vec p_3) \cr
&=& {1\over{2}} (\vec p_2 - \vec p_3 - 
\vec p_{23} {{E_2 - E_3} \over{E_{23} + \sqrt{E_{23}^2 - 
\vec p_{23}^{~2}}  } }) ,
\label{eq_n2}
\end{eqnarray}
with $E_i = \sqrt{m^2 + {\vec p}_i^{\ 2}}$, $E_{23} = E_2 + E_3$ and
 $\vec p_{23} = \vec p_2 + \vec p_3$. 
This is a relativistic generalization of the
nonrelativistic relative momentum $\vec p$. 

The individual momentum state with momentum $ \vec k$ 
for an on-the-mass-shell particle with mass $m$
is defined in terms of the 
state in the rest frame as ~\cite{kei91,Weinberg}
\begin{eqnarray}
U(\beta(k) ) \vert \vec 0 \mu> &=& 
 \sqrt{ {\omega_m({\vec k})  \over{ m }}  }
 \vert \vec k \mu> .
\label{eqn3}
\end{eqnarray}
Here $U(\beta(k))$ is the unitary operator 
 related to the special (along $\vec k$) boost matrix
 $\beta(k)$, $\mu$ the spin magnetic 
quantum number and $ k= ( \omega_m({\vec k}), {\vec k}) $
with
 $\omega_m({\vec k}) = \sqrt{m^2 + {\vec k}^{\ 2}}$. 
Note that by definition $\mu$ does not change. The $4\times 4$ matrix 
$\beta(k)$ is given 
in the Appendix.  

A following unitary operation related to the general boost $\Lambda$ leads to 
the well known Wigner rotation ~\cite{kei91,Weinberg}: 
\begin{eqnarray}
U(\Lambda) \vert \vec k \mu> &=& 
 \sqrt{ { {m}  \over{ \omega_m({\vec k}) } } } 
 U(\Lambda) U( \beta(k) ) 
 \vert \vec 0 \mu> \cr 
&=& U( \beta(k') ) U( \beta^{-1}(k')  \Lambda \beta(k) ) 
 \vert \vec 0 \mu> 
 \sqrt{ { {m}  \over{ \omega_m({\vec k}) } } }  .
\label{eqn4}
\end{eqnarray}

The argument in the second unitary operator is a rotation matrix
\begin{eqnarray}
R(\Lambda, {\vec k}) &\equiv& \beta^{-1}(k') \Lambda \beta(k)
\label{eqn5}
\end{eqnarray}
and often denoted as Wigner rotation. Further the on shell momenta             
    $k$ and $k '$ are related by $k'= \Lambda k$.

$R(\Lambda, {\vec k})$ being a rotation yields
\begin{eqnarray}
U(\Lambda) \vert \vec k \mu> &=& U( \beta(k') ) 
\sum_{\mu ' }  D^{{1\over{2}}}_{\mu' \mu} (R(\Lambda, {\vec k}) ) 
 \vert \vec 0 \mu'>
 \sqrt{ { {m}  \over{ \omega_m({\vec k}) }}  }  ,
\label{eqn6}
\end{eqnarray}
and using (\ref{eqn3}) again to
\begin{eqnarray}
U(\Lambda) \vert \vec k \mu> &=& 
 \sqrt{ {  \omega_m({\vec k}^{\ \prime}) \over{ \omega_m({\vec k}) }}  } 
\sum_{\mu^\prime}  D^{{1\over{2}}}_{\mu' \mu} (R(\Lambda, {\vec k}) ) 
 \vert \vec k~' \mu'> .
\label{eqn7}
\end{eqnarray}
Here $D^{{1\over{2}}}_{\mu' \mu}$   are the standard SU(2)  
Wigner D-matrices~\cite{rose} and their arguments Euler angles 
which are related to $R$ as shown below. 

Now we apply a  general boost to the noninteracting  two nucleon state 
\begin{eqnarray}
\vert \vec k; \vec 0  \mu_2 \mu_3> &\equiv& 
\vert \vec k \mu_2, -\vec k \mu_3 > .
\label{eqn8}
\end{eqnarray}

In the notation to the left we changed from the two individual momenta
to the relative momentum $\vec k$ and 
the total two-nucleon momentum zero. 
 Since the two nucleons are noninteracting $U(\Lambda)$ acting on a two-body 
system is a tensor product
\begin{eqnarray}
U(\Lambda) &=& U_2(\Lambda)  U_3(\Lambda)
\label{eqn9}
\end{eqnarray}
acting on the two spaces of nucleons 2 and 3.
We choose $\Lambda$ as $\beta(P)$, where
 $P=(P_0, \vec P) \equiv 
(\omega_m({\vec p}_2) + \omega_m({\vec p}_3), 
\vec p_2 + \vec p_3)$. 
That boost matrix $\beta(P)$ maps $\vec k$ into 
$\vec p_2$ and $-\vec k$ into $\vec p_3$.
We obtain 
\begin{eqnarray}
U(\beta(P)) \vert \vec k; \vec 0  \mu_2 \mu_3 > \equiv
U_2(\beta(P)) \vert \vec k \mu_2 > 
U_3(\beta(P)) \vert -\vec k \mu_3 > \cr
 \sqrt{ {  \omega_m({\vec p}_2) \over{ \omega_m({\vec k}) } } } 
\sum_{\mu_2'} 
D^{{1\over{2}}}_{\mu_2' \mu_2} (R(\beta(P), \vec k) ) 
 \vert \vec p_2 \mu_2'> 
 \sqrt{ {  \omega_m({\vec p}_3) \over{ \omega_m({\vec k}) } } } 
\sum_{\mu_3'}  D^{{1\over{2}}}_{\mu_3' \mu_3} 
(R(\beta(P), -\vec k) ) 
 \vert \vec p_3 \mu_3'> .
\label{eqn10}
\end{eqnarray}
Of course $\vec p_2$ and $\vec p_3$ are 
given by the inverse relation to (10) and the
related expression for $-\vec k$. 

On the other hand the noninteracting two-nucleon system with total
momentum  zero can be considered as one object with a mass 
$M_0 = 2\omega_m({\vec k})$ 
 and therefore according to the general relation (\ref{eqn3}) one obtains
\begin{eqnarray}
U(\beta(P)) \vert \vec k; \vec 0  \mu_2 \mu_3> &=& 
 \sqrt{ {  \omega_{M_0}(\vec P) \over{ M_0} }  } 
 \vert \vec k; \vec P  \mu_2 \mu_3 > ,
\label{eqn11}
\end{eqnarray}
and we end up with
\begin{eqnarray}
 \vert \vec k; \vec P  \mu_2 \mu_3>  &=& 
\vert~ { {\partial(\vec p_2~\vec p_3)} 
\over { \partial(\vec P~\vec k) } } 
~ \vert^{ 1\over{2} } 
\sum_{\mu_2'}  D^{{1\over{2}}}_{\mu_2' \mu_2} 
(R(\beta(P), \vec k) ) \cr 
 &\times& \sum_{\mu_3'}
  D^{{1\over{2}}}_{\mu_3' \mu_3} 
(R(\beta(P), -\vec k) ) 
 \vert \vec p_2 \mu_2' \vec p_3 \mu_3' > ,
\label{eqn12}
\end{eqnarray}
where 
\begin{eqnarray}
 \vert~ { {\partial(\vec p_2~\vec p_2)} 
\over { \partial(\vec P~\vec k) } } ~ \vert &=& 
 {M_0 \over{\omega_{M_0}(\vec P) }} ~ 
 {\omega_m(\vec p_2) \over{\omega_{m}(\vec k) }} ~
 {\omega_m(\vec p_3) \over{\omega_{m}(\vec k) }}  \equiv 
N^2 ({\vec p}_2, {\vec p}_3) 
\label{eqn13}
\end{eqnarray}
is the Jacobian for the Lorentz transformation from 
$(\vec p_2  , \vec p_3)$ 
to $(\vec P , \vec k)$.
 
This relation generalizes the one used in \cite{relform} and \cite{kam2002} 
from the spinless case 
to the one with spin. In the nonrelativistic 
case the D-matrices reduce to Kronecker 
symbols and the Jacobian is one. 
Thus $\vert \vec p_2 \vec p_3 \mu_2 \mu_3 >$ 
 equals directly
$\vert \vec k \vec P \mu_2 \mu_3 >$, where 
$ \vec k$ equals the nonrelativistic relative momentum. 

The next step is the transition to partial waves. Firstly one defines the
two-body orbital angular momentum states
\begin{eqnarray}
\vert k~l; \vec 0 \mu_l  \mu_2 \mu_3 > &\equiv& 
\int d\hat k Y^l_{\mu_l}(\hat k) \vert \vec k; 
\vec 0 \mu_2 \mu_3 > .
\label{eqn14}
\end{eqnarray}

Then we couple with the total spin $s$ in the two-nucleon c.m. system
to the total angular momentum $j$ and its magnetic quantum number $\mu$:
\begin{eqnarray}
\vert (ls)jk \mu; \vec 0> &\equiv& \sum_{\mu_2 \mu_3} 
\sum_{\mu_s \mu_l} 
 ( {1\over{2}} \mu_2 {1\over{2}} \mu_3 \vert s \mu_s ) 
 (  l  \mu_l s \mu_s \vert j \mu ) 
 \vert k l \mu_l  \mu_2 \mu_3; \vec 0> .
\label{eqn15}
\end{eqnarray}

The special boost $\beta(P)$ leads then to
\begin{eqnarray}
U(\beta(P)) \vert (ls)jk \mu; \vec 0> &=& 
 \sqrt{ {  \omega_{M_0}(\vec P) \over{ M_0} }  } 
\vert (ls)jk \mu; \vec P>
\label{eqn16}
\end{eqnarray}
and applied individually leads finally to
\begin{eqnarray}
\vert (ls)jk \mu; \vec P> &=& 
\vert~ { {\partial(\vec p_2~\vec p_3)} 
\over { \partial(\vec P~\vec k) } } 
~ \vert^{ 1\over{2} } 
\sum_{\mu_2 \mu_3} \sum_{\mu_s \mu_l}  
( {1\over{2}} \mu_2 {1\over{2}} \mu_3 \vert s \mu_s ) 
 (  l  \mu_l s \mu_s \vert j \mu ) 
 \int d\hat k Y^l_{\mu_l}(\hat k) \cr
 &\times& \sum_{\mu_2'}  D^{{1\over{2}}}_{\mu_2' \mu_2} 
(R(\beta(P), \vec k) ) 
\sum_{\mu_3'}
  D^{{1\over{2}}}_{\mu_3' \mu_3} 
(R(\beta(P), -\vec k) ) 
 \vert \vec p_2 \mu_2' \vec p_3 \mu_3'> .
\label{eqn17}
\end{eqnarray}
This is the connection  of the partial wave projected two-nucleon state
with internal momentum k and total momentum $\vec P$ 
to arbitrary individual momentum and spin states.

%

Another requisite is the determination of the Euler angles 
($\alpha, \beta, \gamma$)  in the spin 
1/2 D-matrices. According to (13) the two $4 \times 4$ matrices
\begin{eqnarray}
R(\beta(P), \vec k) &=& \beta^{-1}(p_2) \beta(P) \beta(\omega_m ({\vec k}),
\vec k) \cr
R(\beta(P), -\vec k) &=& \beta^{-1}(p_3) \beta(P) \beta(\omega_m ({\vec k}),
 -\vec k)
\label{eqm1}
\end{eqnarray}
representing rotations 
have the structure $$\left(\matrix{1&0\cr0&M\cr}\right) ,$$
where M is the unitary $3\times 3$ matrix for the Wigner rotation. It
has generally the form ~\cite{rose} 
\begin{eqnarray}
M &=& \left(\matrix{ \cos \alpha \cos \beta \cos \gamma 
- \sin \alpha \sin \gamma &
 \sin \alpha \cos \beta \cos \gamma 
+ \cos \alpha \sin \gamma &
- \sin \beta \cos \gamma \cr
-\cos \alpha \cos \beta \sin \gamma 
- \sin \alpha \cos \gamma &
- \sin \alpha \cos \beta \sin \gamma 
+ \cos \alpha \cos \gamma &
 \sin \beta \sin \gamma \cr
 \cos \alpha \sin \beta  &
  \sin \alpha \sin \beta  &
 \cos \beta  \cr}
\right) .
\label{eqm2}
\end{eqnarray}
This determines the 3 Euler angles. 
The matrix $M$ related to the first equation in (\ref{eqm1}) 
is given in the Appendix. 

It remains to add the third free particle whose momentum 
$\vec p_1$  together with 
the total two-nucleon momentum ${\vec P}$ adds up 
to zero in the 3N c.m. system.

Sticking to our standard nonrelativistic 
notation we denote the orbital angular momentum
of that third particle by $\lambda$ and couple it 
with its spin to its total angular momentum
$I$. Then the 3N partial wave state is
\begin{eqnarray}
\vert k q=p_1 \alpha > &\equiv& 
\vert k p_1  (ls)j (\lambda {1\over{2}})I (jI) JM > \vert (t{1\over{2}})TM_T>
 \cr
 &=& N(\vec p_2, \vec p_3) 
 \sum_{\mu_2 \mu_3 \mu_s }
 \sum_{\mu_l {\mu'}_2 {\mu'}_3} 
\sum_{\mu_1 \mu_{\lambda} \mu_I \mu}  
 ( {1\over{2}} \mu_2 {1\over{2}} \mu_3 \vert s \mu_s ) 
 (  l  \mu_l s \mu_s \vert j \mu ) 
(\lambda \mu_{\lambda} {1\over{2}} \mu_1 \vert I \mu_I )
( j \mu I \mu_I \vert J M) \cr
 &~& \int d\hat p_1 Y^{\lambda}_{\mu_{\lambda}}(\hat p_1) 
\int d\hat k Y^l_{\mu_l}(\hat k) 
D^{1\over{2} }_{{\mu'}_2 \mu_2}(R(\beta(P), \vec k)) 
~D^{1\over{2} }_{{\mu'}_3 \mu_3}(R(\beta(P), -\vec k)) ~ 
 \cr 
 &~& \vert  \vec p_2 {{\mu'}_2}   
\vec p_3 {\mu'}_3 \vec p_1 \mu_1  > .
\label{eqm3}
\end{eqnarray}
In that expression $\vec p_2$ and $\vec p_3$ 
 are functions of $\vec k$ and 
$\vec q = -\vec p_1$.

For the evaluation  of the partial wave representation of
the permutation operator P we need 
the projection of that
state $\vert k q=p_1 \alpha>$  onto 
$<  \vec p_1~' \mu_1' \vec p_2~' \mu_2' \vec p_3~' \mu_3' \vert $. 
Doing that one encounters
($ {\vec P} = - {\vec p}_1 $)
\begin{eqnarray}
&~&\delta( \vec p_2~' - \vec p_2(\vec k, 
\vec P) ) 
\delta( \vec p_3~' - \vec p_3(\vec k, 
\vec P) )  \cr
&=& { 1 \over { \vert~ { {\partial(\vec p_2~' , ~\vec p_3~')} 
\over { \partial(\vec k , ~\vec P) } } 
~ \vert } } 
\delta( \vec k - \vec k(\vec p_2~', 
\vec p_3~') ) 
\delta( \vec P - \vec p_2~' - \vec p_3~')  
\label{eqm4}
\end{eqnarray}
This is verified for instance by integrating 
both sides over $\vec  p_2~'$ and $\vec p_3~'$ and
by converting the integral on the 
right hand side to an integral over $\vec  k$ and 
$\vec P$.
 Thus using (28) one obtains 
\begin{eqnarray}
&~& 
< \vec p_1 m_1  \vec p_2 m_2 \vec p_3 m_3  
 \vert k~ q~ \alpha > ~
= ~ \delta( \vec p_1 + \vec p_2 
+ \vec p_3 ) 
{1\over{ N(\vec p_2, \vec p_3) } } \cr
&~& { \delta(q -p_1) \over { q~p_1 } } ~
{ \delta(k -k(~ \vec p_2,\vec p_3~ ) ) 
\over { k~k(~ \vec p_2,\vec p_3 )~ ) } } ~ \cr
&~& 
 \sum_{\mu_2 \mu_3 \mu_s }
 \sum_{\mu_l \mu_{\lambda} \mu_I \mu}
  ( {1\over{2}} \mu_2 {1\over{2}} \mu_3 \vert s \mu_s ) 
 (  l  \mu_l s \mu_s \vert j \mu ) 
(\lambda \mu_{\lambda} {1\over{2}} m_1 \vert I \mu_I )
( j \mu I \mu_I \vert J M) 
 Y^{\lambda}_{\mu_{\lambda}}(\hat p_1~) 
  Y^l_{\mu_l}(\hat k(~ \vec p_2,\vec p_3 )~ )  \cr
&~& 
D^{1\over{2} }_{m_2 \mu_2}
(R(\beta(P(~ \vec p_2,\vec p_3 )~ ), 
\vec k(~ \vec p_2,\vec p_3 )~ )) \cr 
 &~& 
D^{1\over{2} }_{m_3 \mu_3}
(R(\beta(P(~ \vec p_2,\vec p_3 )~ ), 
-\vec k(~ \vec p_2,\vec p_3 )~ )) .
\label{eqm5}
\end{eqnarray}

This is the basic expression needed for 
the evaluation of the partial wave representation of
the permutation operator P. 
Equipped with that, projecting  Eq.(1) onto the basis states
(28) one encounters like in the nonrelativistic notation ~\cite{book} 
\begin{eqnarray}
_1< k q \alpha \vert P \vert k' q' \alpha' >_1  
&=&  _1< k q \alpha \vert k' q' \alpha' >_2 + 
  _1< k q \alpha \vert k' q' \alpha' >_3 
 =  2~  _1< k q \alpha \vert k' q' \alpha' >_2 .
\label{eqm6}
\end{eqnarray}
This is evaluated  by inserting the complete basis of states
 $\vert \vec  p_1 \mu_1 \vec  p_2 \mu_2 \vec p_3 \mu_3  >$ and using (30).
 The result is worked out in the Appendix. It can be expressed in a form
 which resembles closely the
one appearing in the nonrelativistic regime ~\cite{book,glo96} 
\begin{eqnarray}
_1< k ~ q ~ \alpha \vert ~ P ~ \vert k' ~ q' ~ \alpha' >_1 &=& 
\int_{-1}^{1} dx { {\delta(k-\pi_1)} \over { k^{l+2} }  } ~ 
{ {\delta(k'-\pi_2)} \over { k'^{l'+2} }  } ~  \cr 
&~& 
{1\over {N_1(q, q',x) } } ~ {1\over {N_2(q, q',x) } } ~ 
G_{\alpha \alpha'} (q, q', x) ,
\label{eqm7}
\end{eqnarray}
where all ingredients are given in the Appendix. 

It remains to regard the free propagator adjacent to 
the permutation operator P in Eq.(\ref{eq1a}).
Since only momenta are involved in the propagator the convenient 
formal steps outlaid in ~\cite{kam00}
can be shown in a momentum vector notation thereby simplifying the 
notation. Let
 $\vert \vec k, \vec q>_1$ denote
the 3N state expressed in vector momenta analogous to (20) and neglecting 
spin and
isospin degrees of freedom. The index 1 indicates as above that the 
2N subsystem (23) has
been chosen which is described by the internal momentum $\vec k$. 
Similarily 
an index 2
indicates the choice of the (31) subsystem. Then one obtains
\begin{eqnarray}
&~& _1< \vec k \vec q~ \vert G_0 P_{12} P_{23} \vert 
 \vec k~' \vec q~' >_1 \equiv 
_1< \vec k \vec q~ \vert G_0 \vert 
 \vec k~' \vec q~' >_2 \cr 
&=& \int \prod d\vec p_i ~ 
 _1< \vec k \vec q~ \vert G_0 \vert 
 \vec p_1 \vec p_2 \vec p_3 > 
 \delta( \vec p_1 + \vec p_2 + \vec p_3 ) 
 < \vec p_1 \vec p_2 \vec p_3 \vert 
 \vec k~' \vec q~' >_2 \cr 
&=& \int \prod d\vec p_i 
 {1 \over {E + i\epsilon - \sqrt{m^2 + \vec p_1^{~2}} 
 - \sqrt{m^2 + \vec p_2^{~2}}
 - \sqrt{m^2 + \vec p_3^{~2}} 
 } } ~  
 _1< \vec k \vec q~ \vert 
 \vec p_1 \vec p_2 \vec p_3 > \cr 
&~& 
 \delta( \vec p_1 + \vec p_2 + \vec p_3 ) 
 < \vec p_1 \vec p_2 \vec p_3 \vert 
 \vec k~' \vec q~' >_2 \cr 
&=& \int \prod d\vec p_i 
 {1 \over {E + i\epsilon - \sqrt{m^2 + \vec p_1^{~2}} 
 - \sqrt{m^2 + \vec p_2^{~2}}
 - \sqrt{m^2 + \vec p_3^{~2}} 
 } } 
 \delta( \vec q - \vec p_1 ) 
 \delta( \vec k - 
 \vec k(\vec p_2, \vec p_3) ) \cr 
&~& {1 \over { N(\vec p_2, \vec p_3) }  } 
 \delta( \vec p_1 + \vec p_2 + \vec p_3 ) 
 \delta( \vec q~' - \vec p_2 ) 
\delta( \vec k~' - 
 \vec k(\vec p_3, \vec p_1) ) 
 {1 \over { N(\vec p_3, \vec p_1) }  } \cr 
&=& 
 {1 \over {E + i\epsilon - \sqrt{m^2 + \vec q^{~2}} 
 - \sqrt{m^2 + \vec q~'^2}
 - \sqrt{m^2 + (\vec q + \vec q~')^2}  
 } } 
\delta( \vec k - 
 \vec k(\vec q~', -\vec q 
-\vec q~') ) \cr 
&~& 
 {1 \over { N(\vec q~', -\vec q - \vec q~') }  }
\delta( \vec k~' - 
 \vec k(-\vec q - \vec q~',  
 \vec q) ) 
 {1 \over { N(-\vec q - \vec q~', \vec q~) }  }
\label{eqm8}
\end{eqnarray}

We recognize that the free propagator depends 
on $q$, $q'$ and $x \equiv {\hat q} \cdot {\hat q}^{\, '}$ like in the
nonrelativistic case. After partial wave 
decomposition  there arises an
integration over the interval $[-1,1]$ for $x$. This leads to 
the well known logarithmic
singularities, which have been well studied for the 
nonrelativistic  free propagator.
 That
nonrelativistic propagator results simply by expanding the square roots in 
Eq.~(\ref{eqm8})
and keeping the leading
terms. Like in the 
nonrelativistic case it is now  convenient to put the free propagator
 into
the form  $ \propto 1/( x_0 -x +i\epsilon)$ . A simple algebra leads 
in obvious notation to
\begin{eqnarray}
{1 \over {E +i\epsilon - E_{\vec q} -E_{{\vec q}~'} -E_{\vec q 
+ \vec q~'} 
 } } &=& { A \over { x_0 - x }} ,
\label{eqm9}
\end{eqnarray}
with
\begin{eqnarray}
A &=& { { E - E_{\vec q} -E_{{\vec q}~'} + 
E_{\vec q + \vec q~'}} \over 
{ 2 q q' } } \cr
x_0 &=& { { (E - E_{\vec q} -E_{{\vec q}~'})^2 - m^2 -q^2 -q'^2 } \over 
{ 2 q q' } } .
\label{eqm10}
\end{eqnarray}

Altogether we end up with the 
infinite system of coupled integral equations analogous to
the one in the nonrelativistic case ~\cite{wit88,glo96}:
\begin{eqnarray}
< k q \alpha \vert T(E) \vert  \phi >  &=& < k q \alpha \vert t P \vert \phi >
 + \sum_{\alpha ~'} \sum_{l_{\bar {\alpha}} } \int_0^{\infty} dq' q'^2 
\int_{-1}^{1} dx  {{<kl_{\alpha} \vert t^{(\alpha)}(E 
- \sqrt{m^2 + q^2}) \vert \pi_1 l_{\bar {\alpha}}  >}
\over
{\pi_1^{l_{\bar {\alpha}}}  }} \cr
&~& \times { 
{ G_{\bar {\alpha} \alpha' }(q, q', x) }
\over {   {N_1(q, q',x) }  ~  {N_2(q, q',x) }  ~  }} 
{ {  < \pi_2 q' \alpha' \vert T(E) \vert  \phi >  }
\over{  {\pi_2}^{l_{\alpha'}} }} \cr
&~& \times {{A }\over{  x_0 + i\epsilon - x }} .
\label{eq1f}
\end{eqnarray}

The geometrical coefficients  $G_{\bar {\alpha} \alpha' }(q, q', x)$, 
the coefficients  $N_1(q, q',x)$ and  $N_2(q, q',x)$,  and
the momenta $\pi_1$ and $\pi_2$ stem
from the matrix element  $<kq \alpha \vert P \vert k'q'\alpha'>$ of the
permutation operator (Eqs.(\ref{eq21}) and (\ref{eq21a})~). 
The quantum numbers in the set $\bar {\alpha}$ differ from those in $\alpha$ 
only in the orbital angular momentum $l$ of the pair. 

As mentioned earlier the main problem of treating Eq.(\ref{eq1f}) is caused by
the singularities of the free propagator $G_0$ which occur in the region
of q and q' values for which $\vert x_0 \vert \le 1$. In addition at $q
= q_0$ there
is the singularity of the 2N t-matrix which 
occurs in the $^3S_1-^3D_1$ partial wave
state, where the deuteron bound state exists. 
 The method to treat this singularity is described in detail in
\cite{wit88}. It amounts to separate all channels $\alpha$ which are 
``deuteron''-like (angular momentum quantum numbers $l=0$ or $2$, $s=1$ and $j=1$) 
from the others. In the ``deuteron''-like channels
one separates the bound state 
pole and treats it by subtraction~\cite{wit88}. Similarly,
the treatment of the $G_0$ singularities follows the nonrelativistic case as
described in detail in ref.~\cite{wit88}. 
The only difference is that the relativistic
boundaries of the q,q'-values at which $\vert x_0 \vert = 1$ differ 
from the nonrelativistic boundaries. 
Equation (\ref{eq1f}) is solved by generating its Neumann
series, which is then summed up by the Pad\'e method. 

Due to  short-range nature of the NN force it can be considered
negligible beyond a certain value $j_{max}$ of the total angular momentum
in the two nucleon subsystem. Generally with increasing energy
$j_{max}$ will also increase. For $j > j_{max}$ we put the t-matrix to be zero,
which yields a finite number of coupled channels for each total angular
momentum J and total parity $\pi=(-)^{l+\lambda}$ of the 3N system. 
To achieve converged results at our energies we used all partial 
wave states with total
angular momenta of the 2N subsystem  up to $j_{max}=5$ and took into account 
 all total
angular momenta of the 3N system up to $J=25/2$. This leads to a system
of up to 143 coupled integral equations in two continuous variables
for a given $J$ and parity.

\section{The boosted potential}

As  dynamical input we used a relativistic interaction $v$, 
which is defined as partner of the relativistic kinetic energy, 
generated from 
the nonrelativistic NN potential AV18
according to the analytical prescription of ref.~\cite{kam98}. For
the convenience of the reader we repeat the main points of this
 transformation. Having
a NN potential $v^{nr}$ which provides a nonrelativistic t-matrix $t^{nr}$
obeying the Lippmann-Schwinger equation with the nonrelativistic form of
the free propagator one can apply an analytical 
transformation of momenta to
obtain  an exactly on-shell equivalent relativistic potential
$v^{rl}$ which provides the corresponding relativistic t-matrix
$t^{rl}$. This t-matrix obeys the Lippmann-Schwinger equation 
with a relativistic form of the free propagator. This analytical 
 transformation for the potentials is~\cite{kam98} 
\begin{eqnarray}
v(\vec k, \vec k~') &=& {1\over{h(k_{nr})}} 
v^{(nr)} (\vec k_{nr}, \vec k~'_{nr}) 
{1\over{h(k'_{nr})}}
\label{anal1}
\end{eqnarray}
where
\begin{eqnarray}
k_{nr} \equiv \mid {\vec k}_{nr} \mid & = & \sqrt{2m} \sqrt{\sqrt{{\vec k}^{\ 2} + m^2} - m} \cr
h(k_{nr}) &=& \sqrt{( 1 + {k_{nr}^2\over{2m^2}}) 
\sqrt{ 1 + {k_{nr}^2\over{4m^2}} } } .
\label{anal2}
\end{eqnarray}

In ref.~\cite{kam2002}  it 
was shown that the explicit calculation of the matrix elements 
$V( \vec k, \vec k~' ;  \vec q~ )$
according to Eq.~(4) for the boosted potential
requires the knowledge of the NN bound state wave function and the
half-shell NN t-matrices in the 2N c.m. system. 
In this first study we do not treat the boosted potential matrix element 
 in all its complexity as given in ref.~\cite{kam2002} but 
restrict ourselves to the leading order term in a $q/\omega$ and 
$v/\omega$ expansion  
\begin{eqnarray}
V(\vec k, \vec k~'; \vec q~)  &=&  
v(\vec k, \vec k~')~ \cr 
&\times&[~ 1  
 -  
{{\vec q}^{\ 2} \over{8\sqrt{m^2 + {\vec k}^{\ 2}}  
\sqrt{m^2 +  ( {\vec k}^{\, \prime} )^2 }}} ~ ] .
\label{eq2apr}
\end{eqnarray}

It is therefore important to 
check the quality of such an approximation. To that aim we calculated
the deuteron wave function $\phi_d(\vec k)$ for the  deuteron 
moving with  momentum $\vec q$ using Eq.(\ref{eq2b}). This 
wave function  depends only on the 2N c.m. relative 
momentum $\vec k$ inside the deuteron 
and is thus independent from  the total momentum 
$\vec q$.

\begin{figure}
\includegraphics[scale=0.5]{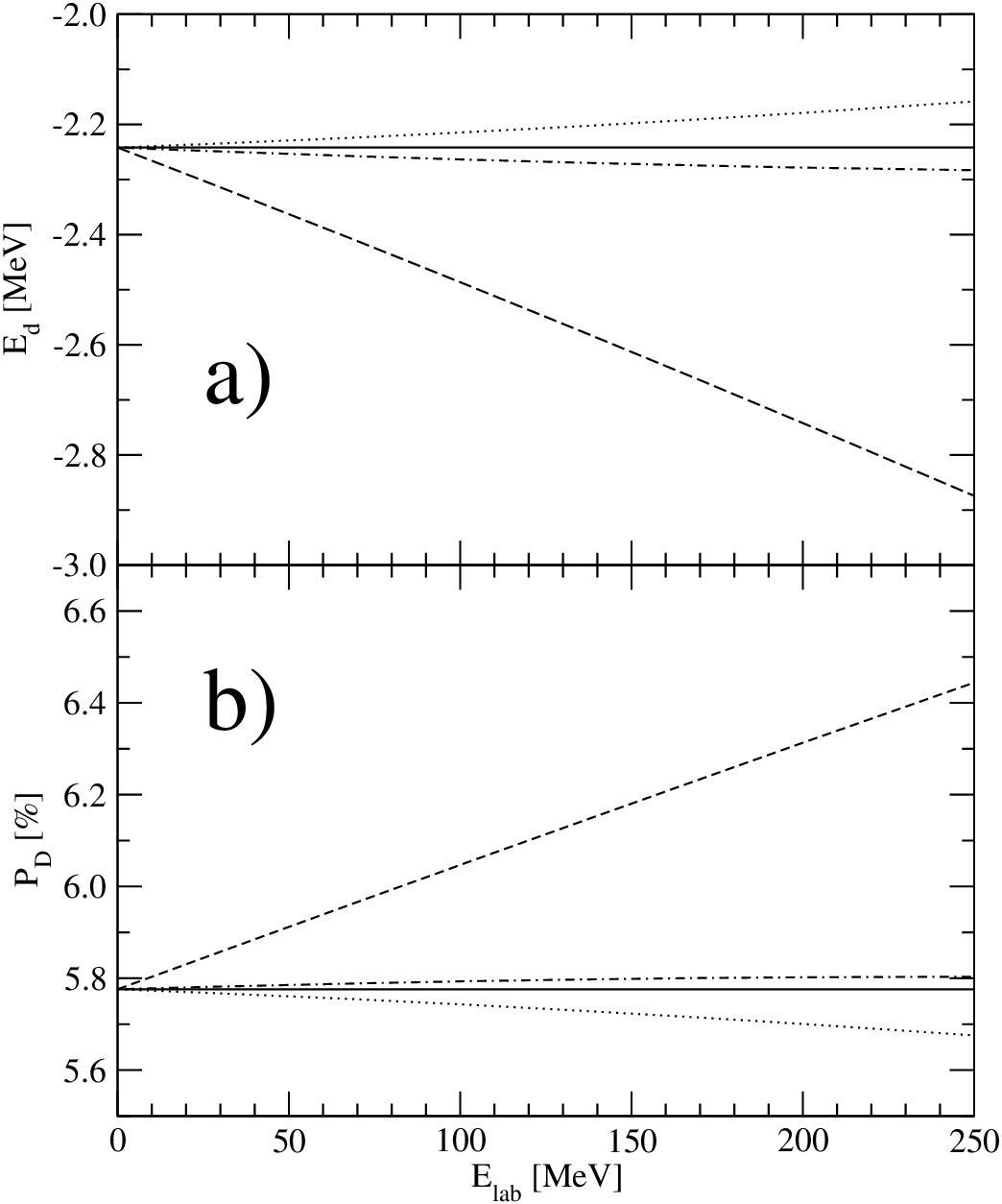}
\caption{The deuteron binding energy $E_d$ and the D-state probability $P_D$. 
The values 
for the deuteron at rest are given by the solid horizontal lines. 
At different incoming nucleon lab. energies,
related to the relative momentum $q_0$ in Eq.~(6),
the approximations given in Eqs.(\ref{ap1}), (\ref{ap2}), and  (\ref{eq2apr}) 
provide results which are shown as the dashed-, dotted-, and dashed-dotted   lines, 
respectively. The calculations have 
been done  with the AV18 potential.
}
\label{figdeuteb}
\end{figure}

We show in Fig.~\ref{figdeuteb}    
the binding energy $E_d$ and the D-state probability $P_D$ 
defined through Eq.~(\ref{eq2b})
as a function of the initial nucleon lab. energy
using the approximation given in Eq.(\ref{eq2apr}). In addition, the
results for two more drastic approximations are given. 
In the first one  the boost effects 
are neglected completely  
\begin{eqnarray}
V(\vec k, \vec k~'; \vec q~)  &=&  
v(\vec k, \vec k~') ,
\label{ap1}
\end{eqnarray}
and in the second one the k-dependence of the first order 
relativistic correction term is 
omitted 
\begin{eqnarray}
V(\vec k, \vec k~'; \vec q~)  &=&  
v(\vec k, \vec k~')~ \left(~ 1 - 
{{\vec q}^{\ 2} \over{8m^2 }} ~ \right) .
\label{ap2}
\end{eqnarray} 

When  the boost effects are fully taken into account the 
solution of Eq.(\ref{eq2b}) must provide exactly the 
deuteron binding energy and the D-state probability equal to the values
for the deuteron at rest. 
It is seen in Fig.\ref{figdeuteb}  
that  neglecting the boost totally or 
omitting the k-dependence of the first order term 
is a poor approximation, especially at the higher energies.
In contrast, the approximation given in Eq.(\ref{eq2apr}) appears acceptable,
even for the strongest boosts.
Relying on that result we have chosen the expression (\ref{eq2apr}) 
for the boosted potential in the following investigations.

\section{Results}


The solution of the 3N relativised Faddeev equation including 
Wigner spin
rotations increase 
the
computer time drastically.
This is caused by  the calculation of the permutation matrix elements 
for the high partial waves.
Therefore to study the effects of the Wigner spin
rotations on the elastic scattering observables 
we restricted ourselves to the $j <  2$ partial
wave states. 
We checked that to get converged results for the permutation matrix
 elements one has to take into account the expansion coefficients 
$a^{\mu_2 \mu_3 \mu_2' \mu_3' m_1 m_2}_{LML'M'}(q,q')$ 
of Eq.(\ref{eq19})  with 
 L,L'  up to L,L'$ \le 2$. Those coefficients  were obtained by
 numerical integrations over the directions $\hat q$ and $\hat q'$ with 23
 gaussian points for the polar and azimuthal angles.
 We found that the changes of the cross sections due to
Wigner spin rotations are small and stay under  $1 \%$. 
 For spin observables  
 these changes are slightly larger but they do not
 exceed $5 \%$ with the exception of angular regions around 
zero crossings and small values of the observables. 
 Thus when performing the fully converged calculations with 
 $j \le 5$ and $J\le 25/2$  we neglected the Wigner spin rotations completely.
This might be different for breakup observables, which deserves another
investigation.

\begin{figure}
\includegraphics[scale=0.6]{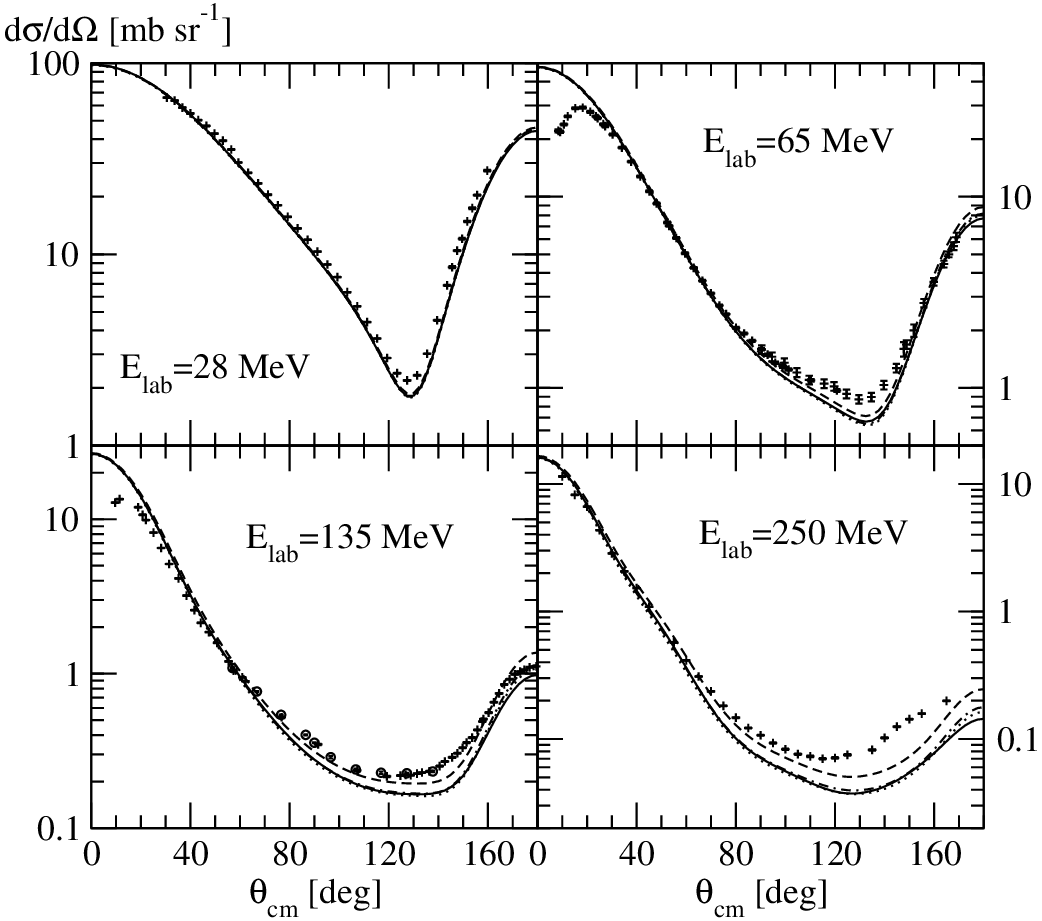}
\caption{
The differential cross sections for  $Nd$ elastic
 scattering at various energies.  The solid
 line is the result of the nonrelativistic Faddeev calculation 
 with the AV18 potential. The relativistic
 predictions based on the approximations 
(\ref{ap1}), (\ref{ap2}), and (\ref{eq2apr})
for the boosted potential keeping all other relativistic features 
unchanged are shown by
 the dashed, dotted, and dashed-dotted  lines.
 The pd  data at $28$, $65$, $135$ and $250$~MeV are from ref.~\cite{hat84}, 
 \cite{shi95},  \cite{sek02}, and \cite{hat02}, respectively. 
}
\label{figsig}
\end{figure}

In Fig.\ref{figsig} we show our results for the nd elastic scattering 
cross sections at four energies together with experimental pd data. In
addition to the nonrelativistic prediction, based on the solution of the 3N
Faddeev equation with the nonrelativistic form of the free propagator $G_0$ and
partial wave states constructed with standard Jacobi momenta, also our relativistic
results using the approximations according to the Eqs.(\ref{ap1}), (\ref{ap2}), 
and (\ref{eq2apr}) 
for the boosted potential matrix elements are presented. 
Thereby the other relativistic features in Eq.~(36) 
have been kept.
It is only the total neglection of the boost effect in $V ({\vec q}\, ) $ which leads 
to a clearly visible deviation from the nonrelativistic results.
Taking the boost effect into account according to the approximations 
(41) and (39) reduces the effect drastically and only at the largest angles 
deviations from the nonrelativistic results are 
discernible.  This is better seen in Fig.~3, where the quantity 
\begin{equation}
\Delta \equiv { {  ( {d\sigma\over {d\Omega}})^{rel} 
 - ( {d\sigma\over {d\Omega}})^{nrel} } \over 
{ ( {d\sigma\over {d\Omega}})^{nrel}  }  }
\label{Delta}
\end{equation}
expressed in percentage
is shown for the three approximations. 
\begin{figure}
\includegraphics[scale=0.6]{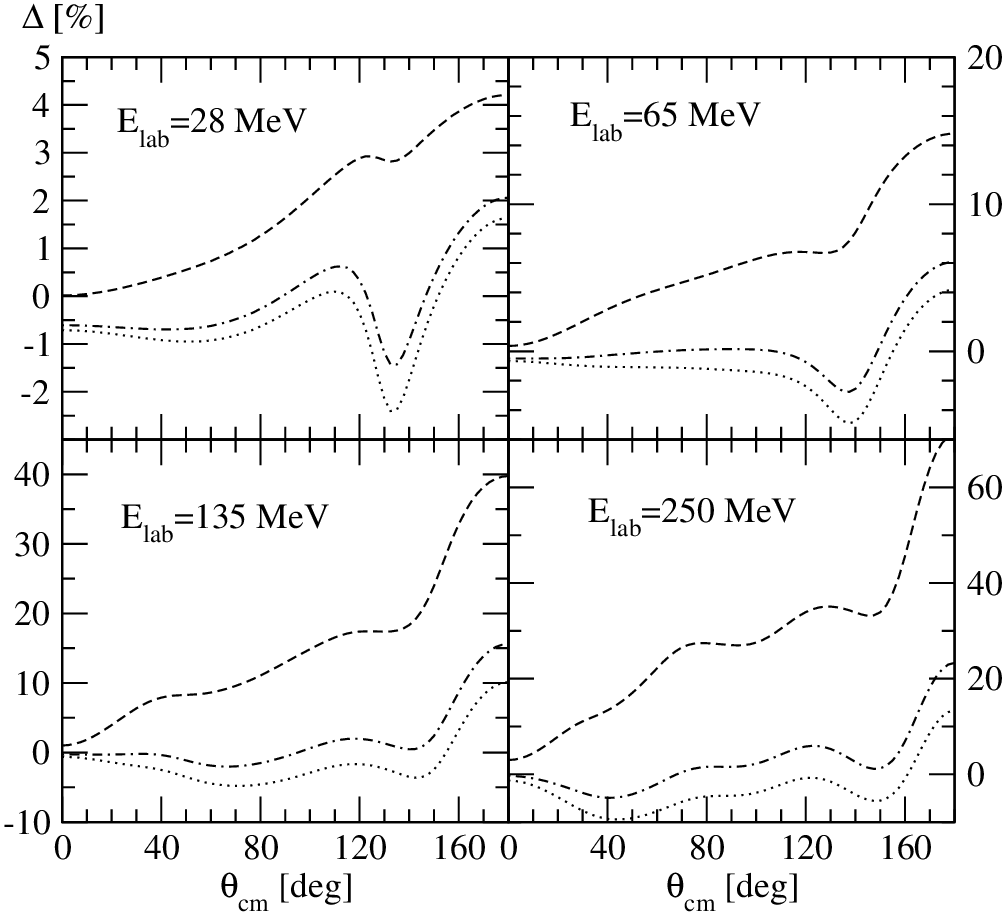}
\caption{The relative deviation $\Delta$ from Eq.~(\ref{Delta})
for the three different relativistic approximations 
to the boosted potential $V( {\vec q} \, )$.
For the description of the lines see Fig.\ref{figsig}.  
}
\label{figdevsig}
\end{figure}
Thus significant effects of
relativity occur at higher energies and they 
are restricted to the backward angle region 
 ($\theta_{c.m} \ge 160^\circ$). 
They increase the nonrelativistic
 cross sections by up to $2 \%$, $6 \%$, $5 - 15 \%$, and 
 $10 - 23 \%$, for 28, 65, 135, and 250 MeV, respectively.  
 For $\theta_{c.m} < 160^\circ$ the effects of relativity
are much smaller. At 250 MeV where they are largest,
they increase the nonrelativistic cross section
 by no more than $\approx 5 \%$
 in the minimum around 
 $\theta_{c.m} = 130^\circ$.
At forward angles the largest effects
 are at $\theta_{c.m} \approx 40^\circ$ where relativity reduces the
 nonrelativistic  cross section by up to $\approx 5 \%$.

In ref.~\cite{wit99} the nd total cross section has been investigated 
in a nonrelativistic scheme.
Beyond that a very first step into relativity has been done 
using the optical theorem. 
In our notation it reads

\begin{eqnarray}
\sigma_{tot} = - {2 \over {\vert \vec j \vert} } 
Im \, \sum\limits_{\mu_n, \mu_d} 
< \phi ~ \mu_n, \mu_d \vert U \vert \phi ~ \mu_n, \mu_d > .
\label{eq02}
\end{eqnarray}

Now we can investigate the changes in both ingredients on 
the right hand side due to
relativity, whereas in \cite{wit99} only the kinematical 
flux quantity $| {\vec j} |$ has been considered.
The ratio of the relativistic to the nonrelativistic flux is given as

\begin{eqnarray}
 { \vert \vec j \vert^{nrel} \over
 { \vert \vec j \vert^{rel} } } =
{ E_n E_d \over { q_0^{rel} (E_n + E_d) } } /
 { m_d m_n \over { q_0^{nrel} (m_d + m_n) } } ,
\label{eq02_a}
\end{eqnarray}
where $E_n$ ($E_d$) is the neutron (deuteron) energy in the c.m. 
system and $q_0^{rel,nrel}$ the
relativistic or nonrelativistic relative momentum in the c.m. system. 
In  \cite{wit99} only
that ratio was considered which led to an increase in the total cross 
section by 3 (7) \%
at 100 (250 ) MeV. Now allowing also for a change of the nuclear matrix element
$  < \phi \vert U \vert \phi  > $ (using the approximate boosted potential 
given in Eq.~(39)) the total cross section (not shown) is
slightly smaller than the nonrelativistic one. In other words, 
the changes in $  < \phi \vert U \vert \phi  > $   outweigh
the kinematical effect, which by itself increases the total cross section. 

Now we come back to the elastic cross section which for the sake 
of completeness and
clarity  is shown.
In our notation it has the form
\begin{eqnarray}
\frac{(d\sigma)^{el,rel}}{ d{\hat q}^{\, \prime} } = 
(2\pi)^4 \left( { E_n E_d \over { E_n + E_d } } \right)^2
\, \frac16 \, \sum\limits_{\mu_n^\prime, \mu_d^\prime, \mu_n, \mu_d} \,
\vert < \phi^\prime ~ \mu_n^\prime, \mu_d^\prime \vert U \vert 
\phi ~ \mu_n, \mu_d > \vert ^2 .
\label{eq02_b}
\end{eqnarray}

The kinematical  factor in the bracket reduces to $\frac{2 m }{3}$  
in the nonrelativistic case. Again we
can regard the ratio of the relativistic and nonrelativistic 
differential  cross sections.
They are  of different type compared to the ratio of the total 
cross sections. Both
ingredients, the kinematical factor and the nuclear matrix element 
enter now squared.
It turned out as we have seen in Figs.~2 and~3 that
 also in this case the dynamical  effects caused by the decrease of 
the nuclear matrix element 
 compensates the increase of the kinematical factor for most angles. 
 Only at very backward angles a slight 
increase remains.
 
\begin{figure}
\includegraphics[scale=0.55]{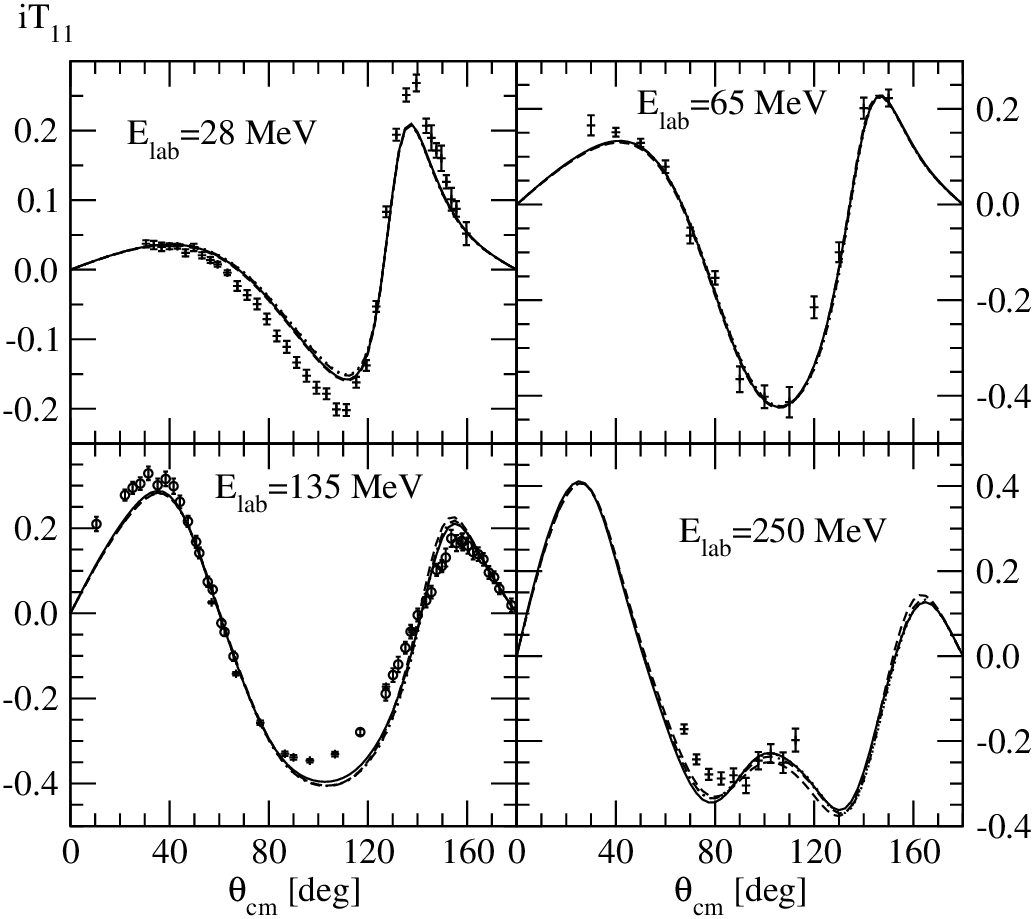}
\caption{
The deuteron vector analyzing power $iT_{11}$ in  $Nd$ elastic
 scattering.  
 For description of lines see Fig.\ref{figsig}. 
 The pd  data at $28$, $65$, $135$ and $250$~MeV are from ref.~\protect\cite{hat84}, 
 \cite{wit93},  \cite{sek02}, and \cite{cad01}, respectively. The pd data 
from \cite{cad01} were taken at $190$~MeV. 
}
\label{figit11}
\end{figure}

\begin{figure}
\includegraphics[scale=0.55]{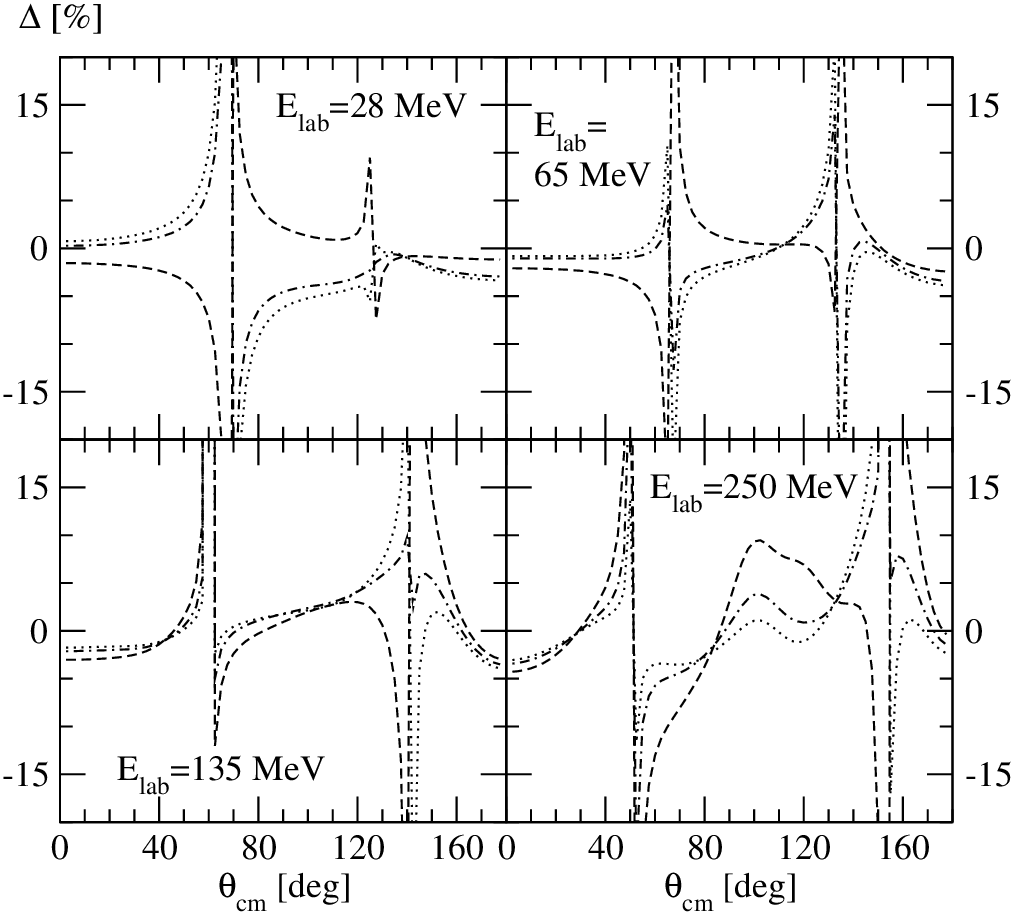}
\caption{
The same as in Fig.\ref{figdevsig} but for the deuteron vector analyzing 
power $iT_{11}$. 
}
\label{figdevit11}
\end{figure}

Finally in Figs.\ref{figit11}-\ref{figdevit11} we compare relativistic and
nonrelativistic predictions for the deuteron vector analyzing 
power $iT_{11}$.  
Again the three approximations to the boosted potential $V ( {\vec q} \, ) $ 
are shown.  It turns out that 
the relativistic effects  are relatively small and 
they stay below $\approx 5 \%$ in the angular regions outside of zero 
crossings.
Other spin observables in elastic scattering behave similarly
and are not shown.

\section{Summary and outlook}

We numerically solved  the 3N Faddeev equation for nd scattering 
including relativistic
features at the  neutron lab energies $E_n^{lab} = 28$, $65$, $135$ 
and $250$~MeV. 
The relativistic features are the relativistic form of the free 
propagator and the change of
the NN potential caused by the boost of the 2N subsystem. In addition 
these boosts
 also induce Wigner
spin rotations. For the momentum space basis we used the relative 
momentum of two  
free  nucleons in their c.m. system together with their total momentum  
which in the 3N
c.m. system is the negative momentum of the 
spectator nucleon. 
Such a choice of momenta is adequate for relativistic kinematics 
and allows to generalize the nonrelativistic approach used to solve the  
nonrelativistic 3N
Faddeev  equation  to the relativistic case in a more or less straightforward 
manner. 
That relative momentum in the two-nucleon subsystem is a generalisation of 
the standard
nonrelativistic Jacobi momentum $\vec p$. 
 The inclusion of the nucleon spins leads 
automatically to Wigner
spin rotations in the context of boosting the 2N c.m. subsystems. The momentum
 partial wave basis, a generalisation of the nonrelativistic one, is, however, 
 now more complex. As dynamical input we took
the nonrelativistic NN potential  AV18  and generated in the 2N c.m. 
system  an exactly on-shell  equivalent relativistic interaction $v$  
 using  an analytical scale transformation of momenta. We checked
 that in our energy range the boost effects for  
 this potential could be sufficiently well incorporated by restricting 
the exact expression 
 to the 
leading order terms in a $q/\omega$ and $v/\omega$  expansion. 

We found that in the studied  energy range the effects of  Wigner spin
rotations are practically negligible for the cross section and analyzing
powers. 
Relativistic effects for the cross section appear at higher energies 
 and they are restricted only to
the very backward angles where relativity increases the nonrelativistic
cross section. At other angles the effects are small. In spite of the
fact that the relativistic phase-space factor increases with energy faster
than the nonrelativistic one, the relativistic nuclear matrix element
outweighs this increase and leads 
for the cross section in a wide angular range  
to a relatively small relativistic effect. 
Also for spin observables (analyzing powers, spin correlation coefficients 
and spin transfer coefficients, not shown) no drastic changes due to relativity
have been found. 

The comparison of our  nonrelativistic theory with existing cross section data
 exhibits at the higher energies clear discrepancies. 
According to our presented results 
  effects due to relativity are significant 
 only in the region of very backward angles where  they 
   increase the cross section. They 
  are relatively small in the region of the cross section
minimum around  $\theta_{cm} \approx 130^{\circ}$, 
where the discrepancies between 
the theory based on pairwise forces only  and data are largest. 
At lower energies (up to about $\approx
 135$~MeV)  this
 discrepancy can be removed when 
 current three-nucleon forces (3NFs), 
 mostly of $2\pi$-exchange character~\cite{TM,uIX}, 
  are included in the nuclear Hamiltonian.  At the higher
 energies, however, a significant part of the discrepancy 
 remains and increases further with increasing energy. 
This indicates  that
additional 3N forces should be added to the $ 2\pi$-exchange type forces. 
Natural candidates in the
traditional meson-exchange picture are exchanges 
like $\pi-\rho$ and $\rho-\rho$. 
This has to be expected since in $\chi$PT \cite{epel2002}  
in the order in which nonvanishing 3NF's appear the first
time there are three topologies of forces, the $2\pi$-exchange, 
a one-pion exchange between one
nucleon and a two-nucleon contact interaction and a pure 3N 
contact interaction. They are
of the same order and have to be kept together. Therefore it appears 
very  worthwhile to
persue a strategy adding in the traditional meson exchange picture 
further 3N forces. Our
results here showing that relativistic effects based on relativistic 
kinematics and
boost effects of the NN force are small support the usefulness of high 
energy elastic Nd
scattering to study 3N force properties.

\section*{Acknowledgments}
This work has been supported by the  Polish Committee for
Scientific Research under Grant no. 2P03B00825, by the NATO grant
no. PST.CLG.978943, and by the Japan Society for the Promotion of
Science. 
H.\ W.\ would like to thank the Triangle Universities Nuclear Laboratory
 and RCNP for hospitality and support during his stay 
in both institutes.
 The numerical
calculations have been performed on the Cray SV1 of the
NIC in J\"ulich, Germany.

\appendix
\section{The special boost matrix $\beta(k)$}

The special Lorentz transformation
$\beta(k) = \beta(~(\omega_m(\vec k ), \vec k)  ~)$ is defined by
$\beta(k) (m, \vec 0) = k$.
It has the following matrix elements:
\begin{eqnarray}
\beta_{ij} &=& \delta_{ij}
+ { { k_i \, k_j} \over { m( m + \omega_m(\vec k )) } }~~~~ (i,j =1,2,3)
\cr
\beta_{\mu 0} &=& \beta_{0 \mu} = { k_{\mu} \over { m } }
~~~~~~~~~~~~~~~ (\mu = 0,1,2,3) .
\label{eq12}
\end{eqnarray}

\section{The matrices of Wigner rotation}

By straightforward calculation, the $ 3 \times 3 $ matrix $M$ 
related to the first equation in Eq.~(\ref{eqm1})
is presented. The four-momentum $P$ is such that
\begin{eqnarray}
\beta(P)~(\omega_m (\vec k), \vec k)~) =
(\omega_m(\vec p_2), \vec p_2)   \cr
\beta(P)~(\omega_m(\vec k), -\vec k)~) =
(\omega_m(\vec p_3), \vec p_3) .
\label{eq14}
\end{eqnarray}
We obtain ($i, j=1, 2, 3$) 
\begin{eqnarray}
M_{i j} =  \delta_{ i j } \ + \
            f_1 \, k_i \, k_j  \ + \
            f_2 \, P_i \, k_j  \ + \
            f_3 \, k_i \, P_j  \ + \
            f_4 \, P_i \, P_j ,
\label{bigM}
\end{eqnarray}
where the four scalar functions $f_1$, $f_2$, $f_3$ and $f_4$ depend 
on the momenta ${\vec k}$ and ${\vec P}$:
\begin{equation}
f_1= \frac{-{E_0} + {M_0}}
     {\left( m + \frac{{M_0}}{2} \right) \,
   \left( {{\vec k} \cdot {\vec P}} + \frac{{E_0}\,{M_0}}{2} +
        m\,{M_0} \right) }
\label{f1new}
\end{equation}
\begin{equation}
f_2=   \frac{2\,\left( 4\,{{\vec k} \cdot {\vec P}} +
       \left( {E_0} + {M_0} \right) \,
      \left( 2\,m + {M_0} \right)  \right) }{\left( {E_0
      } + {M_0} \right) \,\left( 2\,m + {M_0} \right) \,
      \left( 2\,{{\vec k} \cdot {\vec P}} +
     \left( {E_0} + 2\,m \right) \,{M_0} \right) }
\label{f2new}
\end{equation}
\begin{equation}
f_3=  \frac{-2}{2\,{{\vec k} \cdot {\vec P}} +
     \left( {E_0} + 2\,m \right) \,{M_0}}
\label{f3new}
\end{equation}
\begin{equation}
f_4= \frac{2\,m - {M_0}}
   {\left( {E_0} + {M_0} \right) \,
     \left( 2\,{{\vec k} \cdot {\vec P}} +
   \left( {E_0} + 2\,m \right) \,{M_0} \right) }
\label{f4new}
\end{equation}
The quantities $M_0$ and $E_0$ are
\begin{equation}
M_0 = 2 \sqrt{ m^2 + {\vec k}^{\ 2} }
\label{M0}
\end{equation}
and
\begin{equation}
  E_0 = \sqrt{ M_0^2 + {\vec P}^{\ 2} } .
\label{E0}
\end{equation}
The matrix $M$ related to the second equation in Eq.~(\ref{eqm1}) 
results by replacing 
${\vec k}$ by $-{\vec k}$.
Clearly for ${\vec P} = 0$ there is no rotation. 
If one expands $M_{ij}$ in terms of the components
$P_i$ up to second order, a straightforward calculation leads to 
\begin{equation}
 \beta =
 \sqrt{ 
 \frac{4 \, k_3^2}{M_0^2 ( 2 m + M_0)^2 } \, {\vec P}^{\, 2} \ - \
 \frac{8 \, k_3 P_3 \, {\vec k}\cdot{\vec P}}{M_0^2 ( 2 m + M_0)} \ + \
 \frac{M_0 - 2 m}{M_0^2 ( 2 m + M_0)} \, P_3^2
      } 
      \ + \ {\cal O}( {\vec P}^{\, 2} )
\label{beta}
\end{equation}

This is useful numerically in the determination of the Euler angles 
$\alpha$, $\beta$ and $\gamma$. 
For $\beta = 0$ only the combination ($\alpha + \gamma$) occurs and
one can put for instance $\alpha = 0$. 

\section{Permutation operator}

Using Eq.~(30) twice for the bra state 
$ _1 < k q \alpha \vert $ and 
the ket state
$ \vert k^\prime q^\prime \alpha^\prime  >_2 $
one gets for the matrix element of the permutation 
operator in our partial wave basis:

\begin{eqnarray}
 &~& _1< k ~ q ~ \alpha \vert ~ P ~ \vert k' ~ q' ~ \alpha' >_1 = 
2~ _1< k ~ q ~ \alpha ~ \vert ~ k' ~ q' ~ \alpha' >_2 = \cr 
&~& 
2~ \sum_{m_1 m_2 m_3 }
\sum_{\mu_2 \mu_3 \mu_s }
 \sum_{\mu_l  \mu_{\lambda} \mu_I \mu} 
\sum_{\mu_2' \mu_3' \mu_{s'} }
 \sum_{\mu_{l'}  \mu_{\lambda'} \mu_{I'} \mu'} ~ \cr 
&~& 
(\lambda \mu_{\lambda} {1\over{2}} m_1 \vert I \mu_I ) ~ 
( j \mu I \mu_I \vert J M)  ~ 
 ( {1\over{2}} \mu_2 {1\over{2}} \mu_3 \vert s \mu_s ) ~ 
 (  l  \mu_l s \mu_s \vert j \mu ) \cr 
&~& 
(\lambda' \mu_{\lambda'} {1\over{2}} m_2 \vert I' \mu_{I'} ) ~ 
( j' \mu' I' \mu_{I'} \vert J M)  ~ 
 ( {1\over{2}} \mu_2' {1\over{2}} \mu_3' \vert s' \mu_{s'} ) ~ 
 (  l'  \mu_{l'} s' \mu_{s'} \vert j' \mu' )  \cr 
&~& 
 \int d\hat q ~d\hat q'~ 
{1\over{ N(\vec q~', -\vec q - \vec q~') } }  ~
{1\over{ N( -\vec q - \vec q~', \vec q~ ) } } 
\cr  
&~& 
{ \delta(~k - 
\vert \vec k(~ \vec q~', -\vec q -
\vec q~'~ ) \vert ~ ) 
\over { k^2 } } ~
{ \delta( ~ k' - 
\vert \vec k(~ -\vec q -
\vec q~', \vec q ~ ) \vert ~ ) 
\over { {k'}^2 } } ~ \cr 
&~& 
 Y^{\lambda~*}_{\mu_{\lambda}}(\hat q) ~  
  Y^{l~*}_{\mu_l}(\hat k(~ \vec q~', -\vec q 
-  \vec q~' )~ ) ~ 
 Y^{\lambda'}_{\mu_{\lambda'}}(\hat q~') ~  
  Y^{l'}_{\mu_{l'}}(\hat k(~ -\vec q 
-\vec q~', \vec q~ )~ ) ~  \cr 
&~& 
D^{{1\over{2}}~*}_{m_2 \mu_2}
(R(\beta(P(~ \vec q~', -\vec q 
-\vec q~'~) ), 
\vec k(~ \vec q~', -\vec q 
-\vec q~' )~ )) ~ \cr 
&~& 
D^{{1\over{2}}~*}_{m_3 \mu_3}
(R(\beta(P(~ \vec q~', -\vec q 
-\vec q~'~) ), 
-\vec k(~ \vec q~', -\vec q 
-\vec q~' )~ )) ~  \cr 
&~& 
D^{{1\over{2}}}_{m_3 \mu_2'}
(R(\beta(P(~ -\vec q 
-\vec q~', \vec q ~) ), 
\vec k(~ -\vec q 
-\vec q~', \vec q~ )~ )) ~  \cr 
&~& 
D^{{1\over{2}}}_{m_1 \mu_3'}
(R(\beta(P(~ -\vec q 
-\vec q~', \vec q ~) ), 
-\vec k(~ -\vec q 
-\vec q~', \vec q~ )~ )) ~ \cr 
&~& 
 _1< (~ (t{1\over{2}}) T ~ \vert ~ (t'{1\over{2}}) T ~ >_2 ,
\label{eq17}
\end{eqnarray}
with 
\begin{eqnarray}
&~& \vec k(~\vec q~' , -\vec q 
-\vec q~')~ \equiv ~ \vec q~' + 
{1\over{2}} \vec q (1 + y_1(q, q',x))  ~ \cr
&~& 
\vec k(~ -\vec q 
-\vec q~', \vec q~ )~ \equiv ~ -\vec q - 
{1\over{2}} \vec q~' (1 + y_2(q, q',x)) .
\label{eq18}
\end{eqnarray}
In Eq.(\ref{eq18}) occur
\begin{eqnarray}
y_1(q, q',x) = 
\frac{  E_{ {\vec q}^{\, \prime}} \, - \, 
E_{ {\vec q} + {\vec q}^{\, \prime}}  }
{
E_{ {\vec q}^{\, \prime}} \, + \, E_{ {\vec q} + {\vec q}^{\, \prime}}  \ + \
\sqrt{ ( E_{ {\vec q}^{\, \prime}} \, + 
\, E_{ {\vec q} + {\vec q}^{\, \prime}} )^2
- {\vec q}^{\ 2} }
}
\label{y1}
\end{eqnarray}
with $x=\hat q \cdot \hat q' $, 
$y_2(q, q',x) = y_1(q', q,x)$ and $ E_{ {\vec q} } \equiv \omega_m (
{\vec q~} )$. 

Expanding the product of the four D-matrices in Eq.(\ref{eq17})
into spherical harmonics
\begin{eqnarray}
\sum_{m_3}  ~
&~& D^{{1\over{2}}~*}_{m_2 \mu_2}
(R(\beta(P(~ \vec q~', -\vec q 
-\vec q~'~) ), 
\vec k(~ \vec q~', -\vec q 
-\vec q~' )~ )) ~ \cr 
&~&
D^{{1\over{2}}~*}_{m_3 \mu_3}
(R(\beta(P(~ \vec q~', -\vec q 
-\vec q~'~) ), 
-\vec k(~ \vec q~', -\vec q 
-\vec q~' )~ )) ~  \cr 
&~& 
D^{{1\over{2}}}_{m_3 \mu_2'}
(R(\beta(P(~ -\vec q 
-\vec q~', \vec q ~) ), 
\vec k(~ -\vec q 
-\vec q~', \vec q~ )~ )) ~  \cr 
&~& 
D^{{1\over{2}}}_{m_1 \mu_3'}
(R(\beta(P(~ -\vec q 
-\vec q~', \vec q ~) ), 
-\vec k(~ -\vec q 
-\vec q~', \vec q~ )~ )) ~ \cr 
&~& = \sum_{LML'M'} a^{\mu_2 \mu_3 \mu_2' \mu_3' m_1 m_2}_{LML'M'}(q,q') ~ 
Y^*_{LM}(\hat q)~ Y_{L'M'}(\hat q~')~ ,
\label{eq19}
\end{eqnarray}
and performing the integrations over $\hat q$ and $\hat q'$ in
Eq.(\ref{eq17}) leads to  
 the following expression for the matrix element of the permutation operator P:
\begin{eqnarray}
_1< k ~ q ~ \alpha \vert ~ P ~ \vert k' ~ q' ~ \alpha' >_1 &=& 
\int_{-1}^{1} dx { {\delta(k-\pi_1)} \over { k^{l+2} }  } ~ 
{ {\delta(k'-\pi_2)} \over { k'^{l'+2} }  } ~  \cr 
&~& 
{1\over {N_1(q, q',x) } } ~ {1\over {N_2(q, q',x) } } ~ 
G_{\alpha \alpha'} (q, q', x) ,
\label{eq20}
\end{eqnarray}
with 
\begin{eqnarray}
G_{\alpha \alpha'} (q, q', x) &=& \sum_k P_k(x)~ \sum_{l_1 + l_2 = l}  
\sum_{l_1' + l_2' = l'}  q^{l_2+l_2'}  q~'^{l_1 + l_1'}  
(1 + y_1)^{l_2} (1 + y_2)^{l_1'}  
g_{\alpha \alpha'}^{k l_1 l_2 l_1' l_2'} (q, q') 
\label{eq21}
\end{eqnarray}
and 
\begin{eqnarray}
\pi_1 &=& \sqrt{q'^2 +{1\over{4}}q^2(1+y_1)^2 + qq'x(1+y_1)}  \cr
\pi_2 &=& \sqrt{q^2 +{1\over{4}}q'^2(1+y_2)^2 + qq'x(1+y_2)} \cr 
N_1(q, q',x) &\equiv&  
N(\vec q~', -\vec q - \vec q~')  \cr
N_2(q, q',x) &\equiv& 
N( -\vec q - \vec q~', \vec q~ ).  
\label{eq21a}
\end{eqnarray}
The geometrical coefficients 
$g_{\alpha \alpha'}^{k l_1 l_2 l_1' l_2'} (q, q')$ 
are given by 
\begin{eqnarray}
g_{\alpha \alpha'}^{k l_1 l_2 l_1' l_2'} (q, q') &=& 
\sum_{\mu_2 \mu_3 \mu_2' \mu_3' }
 \sum_{L M L' M'} \sum_{m_1 m_2} 
a^{\mu_2 \mu_3 \mu_2' \mu_3' m_1 m_2}_{LML'M'}(q,q') ~ 
A^{\mu_2 \mu_3 \mu_2' \mu_3' m_1 m_2}_{LML'M'}(J^{\pi} \alpha 
\alpha' k l_1 l_2 l_1' l_2') ,
\label{eq22}
\end{eqnarray}
with 
\begin{eqnarray}
A^{\mu_2 \mu_3 \mu_2' \mu_3' m_1 m_2}_{LML'M'}&(&J^{\pi} \alpha 
\alpha' k l_1 l_2 l_1' l_2') = {1\over {4\pi}} (-1)^{t'} \delta_{TT'} 
\delta_{M_T M_{T'}} \sqrt{\hat t \hat t'} \left\{\matrix{ 
1/2 & 1/2 & t \cr
1/2 &  T  & t' \cr}
\right\} \cr
&~& 
(-1)^k \hat k ({1\over{2}})^{l_2} ({1\over{2}})^{l_1'} (-1)^{l'} 
\sqrt{\hat \lambda \hat \lambda'} 
\sqrt{ {  {(\hat l')! } \over {(2l_1')! (2l_2')!  }  }  }  
\sqrt{ {  {(\hat l)! } \over {(\hat l_1)! (\hat l_2)!  }  }  } 
 { {\sqrt{\hat L \hat L'  }  }\over { \sqrt{ \hat l_2 \hat l_1  }  }  } \cr 
&~& 
( {1\over{2}} \mu_2 {1\over{2}} \mu_3 \vert s \mu_2 + \mu_3 ) ~ 
( {1\over{2}} \mu_2' {1\over{2}} \mu_3' \vert s' \mu_2' + \mu_3' )~ 
\sum_{c_1} (L 0 \lambda 0  \vert c_1 0 ) ~ 
\sum_{c_1'} (L' 0 \lambda' 0  \vert c_1' 0 )~ \cr  
&~& 
\sum_{\mu_{\lambda}} (L M \lambda \mu_{\lambda} \vert c_1 M + \mu_{\lambda} )~
(-1)^{ M + \mu_{\lambda} } 
(\lambda \mu_{\lambda} {1\over{2}} m_1 \vert I \mu_{\lambda} +  m_1 ) ~ \cr 
&~& 
\sum_{\mu} ( j \mu I \mu_{\lambda} +  m_1 \vert J  \mu + \mu_{\lambda} +m_1 ) ~
 (  l  \mu - \mu_2 -\mu_3  s \mu_2 + \mu_3  \vert j \mu ) ~ \cr
&~& 
\sum_{\mu'} 
( j' \mu' I' \mu + \mu_{\lambda} +  m_1 - \mu' \vert J  \mu + 
\mu_{\lambda} +m_1 ) ~
 (  l'  \mu' - \mu_2' -\mu_3'  s' \mu_2' + \mu_3'  \vert j' \mu' ) ~ \cr 
&~& 
(\lambda' \mu + \mu_{\lambda} + m_1 - \mu' -m_2  {1\over{2}} m_2 \vert 
I' \mu + \mu_{\lambda} + m_1 -\mu' ) ~ \cr
&~& 
( L' M'  \lambda' \mu + \mu_{\lambda} + m_1 - \mu' -m_2 \vert 
c_1' \mu + \mu_{\lambda} + m_1 - \mu' -m_2 + M' ) ~ \cr 
&~& 
\sum_{f_1} (k 0 l_1' 0 \vert f_1 0) \sqrt{\hat f_1} 
(c_1' 0 f_1 0 \vert l_1 0) ~ 
\sum_{f_2} (k 0 l_2' 0 \vert f_2 0) \sqrt{\hat f_2} 
(c_1 0 f_2 0 \vert l_2 0) ~ \left\{\matrix{ 
f_2 & f_1 & l' \cr
l_1' &  l_2'  & k \cr}
\right\} ~ \cr 
&~& 
\sum_{m_{l_1}} (l_1 m_{l_1} l_2 \mu - \mu_2 - \mu_3 - m_{l_1} \vert  
l \mu - \mu_2 - \mu_3) ~ \cr 
&~& 
(c_1 -M -\mu_{\lambda} f_2 \mu - \mu_2 - \mu_3 -  m_{l_1} + M + \mu_{\lambda} 
\vert l_2  \mu - \mu_2 - \mu_3 -  m_{l_1} ) ~ \cr 
&~& 
(f_1 m_{f_1} f_2 m_{f_2} \vert l'  \mu' - \mu_2' - \mu_3') ~ \cr 
&~& 
(c_1' \mu + \mu_{\lambda} + m_1 - \mu' -m_2 + M' f_1  
m_{f_1} 
\vert l_1 
m_{l_1}) .
\label{eq23}
\end{eqnarray}
We used our standard notation $ {\hat l} \equiv 2 \, l + 1 $.
The coefficients 
$A^{\mu_2 \mu_3 \mu_2' \mu_3' m_1 m_2}_{LML'M'}(J^{\pi} \alpha 
\alpha' k l_1 l_2 l_1' l_2')$ are real and obey the following 
symmetry property
\begin{eqnarray}
A^{-\mu_2 -\mu_3 -\mu_2' -\mu_3' -m_1 -m_2}_{L-ML'-M'}(J^{\pi} \alpha 
\alpha' k l_1 l_2 l_1' l_2') = 
A^{\mu_2 \mu_3 \mu_2' \mu_3' m_1 m_2}_{LML'M'}(J^{\pi} \alpha 
\alpha' k l_1 l_2 l_1' l_2') .
\label{eq24}
\end{eqnarray}
The proof involves in addition to the definition (\ref{eq23}) 
the properties $l_1 + l_2 = l$, $l_1' + l_2' = l'$
and the parity conservation: 
$ (-1)^{l + \lambda} =  (-1)^{l' + \lambda '}$.
The coefficients $a^{\mu_2 \mu_3 \mu_2' \mu_3' m_1 m_2}_{LML'M'}(q,q')$,
defined in (\ref{eq19}),  
are complex and obey 
\begin{eqnarray}
a^{* \mu_2 \mu_3 \mu_2' \mu_3' m_1 m_2}_{LML'M'}(q,q') =  
(-1)^{M + M' + m _2 - \mu_2 -\mu_3 -\mu_2'+ m_1  -\mu_3' +1 }~
a^{-\mu_2 -\mu_3 -\mu_2' -\mu_3' -m_1 -m_2}_{L-ML'-M'}(q,q') .
\label{eq25}
\end{eqnarray}
To obtain (\ref{eq25}) one uses the relation 
$D^{j\ *}_{m m'} ( \alpha, \beta, \gamma ) 
= (-1)^{m - m'} \, D^{j}_{-m -m'} ( \alpha, \beta, \gamma ) $.
The two last Clebsch-Gordan coefficients 
in Eq.~(\ref{eq23}) provide the condition that the phase factor 
in  Eq.~(\ref{eq25}) is one. 
Thus the geometrical coefficients  
$g_{\alpha \alpha'}^{k l_1 l_2 l_1' l_2'} (q, q')$ are real.

\end{document}